# The microscopic origins of stretched exponential relaxation in two model glass-forming liquids as probed by simulations in the isoconfigurational ensemble

Daniel Diaz Vela[a] and David S. Simmons[a*]

[a] *Department of Chemical and Biomedical Engineering, University of South Florida, Tampa, FL 33620.*

**Abstract:** The origin of stretched exponential relaxation in supercooled glass-forming liquids is one of the central questions regarding the anomalous dynamics of these fluids. The dominant explanation for this phenomenon has long been the proposition that spatial averaging over a heterogeneous distribution of locally exponential relaxation processes leads to stretching. Here we perform simulations of model polymeric and small-molecule glass-formers in the isoconfigurational ensemble to show that stretching instead emerges from a combination of spatial averaging and locally nonexponential relaxation. Results indicate that localities in the fluid exhibiting faster-than-average relaxation tend to exhibit locally stretched relaxation, whereas slower-than-average relaxing domains exhibit compressed exponential relaxation. We show that local stretching is predicted by loose local caging, as measured by the Debye-Waller factor, and vice versa. This phenomenology in the local relaxation of in-equilibrium glasses parallels the dynamics of out of equilibrium under-dense and over-dense glasses, which likewise exhibit an asymmetry in their degree of stretching vs compression. On the basis of these results, we hypothesize that local stretching and compression in equilibrium glass-forming liquids results from evolution of particle mobilities over a single local relaxation time, with slower particles tending towards acceleration and vice versa. In addition to providing new insight into the origins of stretched relaxation, these results have implications for the interpretation of stretching exponents as measured via metrologies such as dielectric spectroscopy: measured stretching exponents cannot universally be interpreted as a direct measure of the breadth of an underlying distribution of relaxation times.

## Introduction

Numerous liquids, ranging from polymers to small organic molecules to metals, apparently solidify via a continuous dynamic arrest process known as the glass transition. Liquids well into this process – in the supercooled liquid regime – exhibit a number of dynamical anomalies that are characteristic of dynamics during glass formation. Perhaps the most prominent of these anomalies is the ubiquitous observation of stretched exponential relaxation – i.e. a structural relaxation process that deviates from a single-time Maxwell process[1]. This type of relaxation is commonly characterized via fits to the Kohlsrauch-Williams-Watts stretched exponential form,

$$F = A\exp\left[-\left(\frac{t}{\tau}\right)^\beta\right], \quad (1)$$

where $\beta$ is the stretching exponent, with a value of 1 for a single-time Maxwell process. Values appreciably less than one are typically observed for glass-forming liquids.

The precise origin of stretched exponential relaxation is one of the core questions in the search for an understanding of the glass transition. The dominant explanation for this phenomenon is the proposition that values of $\beta$ less than one reflect an average over a spatially heterogeneous distribution of local relaxation times[2,3]. Spatial "dynamical heterogeneity" has been directly observed in both simulations[4–7] and colloidal experiments[6,8]. A range of experimental measurements are also consistent with the presence of a distribution of local relaxation times in glass-forming liquids[2,9]. The presence of dynamic heterogeneity in glass-forming liquids is thus now well established. At the same time, it remains unclear whether averaging over spatial heterogeneity is generally the only source of stretching.

An alternate vein of thought suggests that dynamics may in fact be locally nonexponential[10,11], in which case local nonexponentiality may contribute to or even dominate stretching in some systems. This type of phenomenon has been referred to as "intrinsic" nonexponentiality[12]. When distinguishing between stretching that emerges from averaging and genuinely intrinsic stretching, however, it is essential to consider to role of observation time scale. Specifically, because ergodicity in supercooled liquids is restored over sufficiently long timescales, even a 'local' measure of relaxation will sample a full distribution of heterogeneous environments over a sufficient observation window. Indeed, if one defines local relaxation this way, then one is assured of observing equal 'local' and 'global' relaxation times. Throughout this discussion, when we refer to "intrinsic" nonexponentiality, we therefore exclude this time-averaging effect. Instead, by this term we we refer to nonexponentiality of local relaxation over a single local relaxation time, without integration over some finite time window.

Prior efforts have sought to provide insight into the relative roles of intrinsic nonexponentiality and spatial averaging in relaxational stretching. For example, a simulation study by Qian et al. attempted to infer a local contribution to nonexponentiality in a small-molecule glass-former by quantifying dynamics in subsequent time intervals and seeking evidence of temporally correlated dynamics[13]. This effort concluded that local contributions to nonexponentiality dominate during the fast picosecond relaxation (which involves ballistic motion and sub-cage exploration) but that spatial averaging dominates at longer times associated with segmental alpha relaxation. However, this work was constrained by the relatively short times accessible to simulation at that time and the need to indirectly infer a local contribution rather than directly measure it.

More recently, Shang et al probed these questions with more modern computational equipment by shearing a glass-forming-liquid while allowing only a sub-region of controlled size to deform non-affinely[14]. This work found that the stretching exponent associated with the resulting mechanical relaxation was not appreciably size dependent and remained appreciable less than one even for the smallest domains probed, suggesting an important role for local non-exponentiality. However, interpretation of these findings is complicated by the potential presence of nanoconfinement effects associated with frozen matrix approaches[15–20], such that it is not clear that the stretching exponent here is truly analogous to that in an unperturbed bulk liquid. Moreover, this approach involves a nontrivial minimum size scale such that even their limiting small-domain measurements potentially already involve significant spatial averaging[14].

Richert has suggested based on a number of disparate experimental studies, including dielectric hole-burning and multi-dimensional NMR, that spatial heterogeneity is the dominant origin of stretching in small-molecule glass-formers but that local contributions may play an important role in the reorientational relaxation of polymers[9]. In general, these experimental studies establish fairly firmly that dynamical heterogeneity is present and thus that spatial averaging must play an important role in stretching. However, they do not fully resolve the question of whether spatial averaging is the *only* contributor to stretching, or whether local nonexponentiality may play a role as well in some systems.

Several additional reasons have emerged to question whether stretched exponential relaxation emerges purely from spatial averaging over a heterogeneous distribution of locally exponential relaxation processes. First, recent work has raised an apparent contradiction of this proposition with a second anomalous aspect of supercooled liquid dynamics: translational/reorientational decoupling. Specifically, many glass-forming liquids also exhibit a decoupling between the temperature dependence of relaxation processes that in simple liquids exhibit a shared temperature dependence. Examples include breakdown of the Stokes-Einstein relation[3,21,22], decoupling between translational and reorientational relaxation, chain normal mode decoupling in polymers[23,24], and ion-segmental relaxation decoupling in ion-containing polymers[25]. The leading explanation for this phenomenon is also spatial dynamic heterogeneity[3], with distinct relaxation processes predicted to report on distinct moments of this underlying distribution. This explanation requires that the distribution of relaxation times must broaden on cooling, which, with the averaging hypothesis for stretching above, would anticipate a progressive suppression in the stretching exponent on cooling. However, a number of dielectric spectroscopy measurements in systems exhibiting pronounced cooling have reported a nearly temperature invariant dielectric stretching exponent[26]. This finding indicates that either the understanding of decoupling or the understanding of stretching must be incorrect or incomplete.

Pursuant to these earlier findings, recent work by our group has probed the relaxation-function dependence of the stretching exponent[23]. This work indicated that stretching exponents for translational and reorientational relaxation can differ considerably for the same system; moreover, the form of the translational stretching exponent was itself found to depend on wavenumber (inverse length scale). How can stretching exponents report in a simple way on the breadth of an underlying distribution of relaxation times if their behavior varies qualitatively between distinct relaxation functions in the same system? If stretching does not purely reflect spatial averaging, what other mechanisms contribute to its emergence in supercooling liquids?

The most direct experimental efforts to answer these questions have employed single molecule reorientation experiments in an effort to probe the distribution of local dynamics in glass-forming liquids[27–30]. These experiments have provided enormous insight into dynamic heterogeneity in experimental systems, directly reporting on a distribution of relaxation times across distinct molecules[28,29]. Perhaps most importantly, these experiments have reported on a distribution of stretching exponents across distinct molecules, with slower-relaxing molecules tending to exhibit less stretching[28].

On their face, this latter result would seem to provide evidence for local nonexponentiality. However, a key limitation of even these cutting edge experiments prevents this interpretation: because of requirements for adequate statistical sampling, even the shortest such experiments involve time integration over many multiples of the mean system relaxation time. Indeed, lower limits of such sampling periods appear to be of order 100 times the median relaxation time[28,30]. As a consequence, the 'local' relaxation times and stretching exponents reported by this methods are not genuinely 'intrinsic local' properties, but instead reflect temporal averaging and some degree of restoration of ergodicity. Indeed, Paeng et al demonstrated that increasing this observation window leads to a narrowing of the distribution of nominally 'local' (although subject to temporal averaging) stretching exponents, consistent with recovery of ergodicity[28]. Put another way, the reported 'local' values temporally average over heterogeneity, which given sufficient time is equivalent to a spatial average over heterogeneity.

Indeed, this temporal averaging effect was proposed as the favored explanation for the observation described above of apparently more stretched relaxation in faster relaxing moelcules[28]. Paeng et al. attempted to account for this by performing experiments spanning a range of integration windows and extrapolating to an instantaneous measurement[28,29]. As expected, they report a progressive suppression in nonexponentiality with shorter observation windows – a natural results of reduced temporal (equivalent to spatial) averaging[28,29]. They suggested that this behavior might reasonably extrapolate to a recovery of exponentiality for short observation windows and that this could be understood as indicating that stretching emerges entirely from spatial and temporal averaging. However, because their data do not access sufficiently short integration times to observe a return to full exponentiality, it is not a clear whether this recovery of exponentiality for intrinsic local (non-temporally-averaged) dynamics genuinely occurs as a general matter. Moreover, this interpretation does not appear to be congruent with their finding that the distribution of local stretching exponents *broadens* with reducing integration time – a lack of intrinsic local exponentiality should require uniformly exponential local relaxation with a short observation window[12].

A final complicating factor concerns potential variations in the mechanism of nonexponentiality in distinct classes of material. Specifically, unlike in small molecules, in polymers, chain connectivity effects are expected to lead to local non-exponentiality in the relaxation of polymer segments[31]. At the same time, polymers also exhibit emergent dynamic heterogeneity upon cooling towards $T_g$[23,31–35], and presumably any physics governing potential non-exponentialities in small molecules may be present in polymers as well.

In order to overcome these challenges and answer the above questions more clearly, here we perform simulations of model small-molecule and polymer liquids in the isoconfigurational ensemble[36]. This ensemble allows calculation of quantities that are normally defined on a system average basis instead on a single-particle level, and is thus well-suited to the study of dynamical heterogeneity[4] and local properties in glass-forming liquids. Most importantly, this approach allows determination of single-particle properties in simulation *entirely without temporal averaging*. As described below, relaxation times and stretching exponents for each particle are based on the first local relaxation period for that particle, and not on a time-integration over multiple relaxation events. This elimination of temporal averaging permits us to probe directly the role of genuinely intrinsic local mobility as opposed to spatial averaging. Throughout the remainder of this paper, when refer to "local" properties from these simulations we thus refer to intrinsic local properties in the absence of any temporal averaging.

Specifically, these simulations allow us to quantify stretching at both a system-mean and single-particle level and *directly* quantify the extent to which overall relaxational stretching can be attributed to local nonexponentiality vs spatial averaging. Results indicate that dynamics are intrinsically locally stretched, even in the absence of temporal averaging, such that stretching of the mean system dynamics results from a combination of spatial averaging and local stretching.

The spatial averaging contribution is rationally connected to an observed distribution of relaxation times that broadens on cooling towards the glass transition. Local dynamics, on the other hand, exhibit a spectrum of stretching behavior ranging from highly stretched to modestly compressed ($\beta > 1$). Results suggest that this behavior may be a local in-equilibrium analogue of the global behavior of non-equilibrium glasses, wherein individual relaxation events are accompanied by systematic alterations in mobility. Findings indicate that this tendency for relaxation of slow regions to enhance mobility and vice versa is significant *even over a single local relaxation time*, such that it leads to intrinsic stretching in a manner that is intimately related to dynamical heterogeneity but cannot be attributed to either spatial or temporal averaging.

## Methods

We simulate two model systems: a modified bead-spring polymer model, and a rigid polydisperse dimer small-molecule model. All non-bonded beads interact via the binary Lennard-Jones (LJ) potential:

$$E_{ij} = 4\varepsilon_{ij}\left[\left(\frac{\sigma_{ij}}{r}\right)^{12} - \left(\frac{\sigma_{ij}}{r}\right)^{6}\right] \quad r < r_{cut}, \quad (2)$$

where $\sigma_{ij}$ and $\varepsilon_{ij}$ set the range and strength of interactions, respectively, between beads of type $i$ and $j$, and $r_{cut}$ is a cutoff distance. For all simulations and interactions here, $\varepsilon_{ij} = 1$ and $r_{cut} = 2.5$.

Polymer chains each consist of 20 beads bonded linearly. Bonds employ the Finitely Extensible Nonlinear Elastic (FENE) potential,

$$E_F = -0.5KR_0^2 \ln\left[1-\left(\frac{r}{R_0}\right)^2\right] + 4\varepsilon_F\left[\left(\frac{\sigma_F}{r}\right)^{12} - \left(\frac{\sigma_F}{r}\right)^{6}\right] + \varepsilon, \quad (3)$$

where the second term, which defines an LJ repulsion between bonded beads, is cut off at its minimum of $2^{1/6}\sigma_F$. We employ values of $K = 30$, $R_0 = 1.3$, $\varepsilon_F = 0.8$, and $\sigma_F = 1.0$, which represents a modest modification to the standard model of Kremer and Grest[37]. The Kremer-Grest family of polymer models has been widely employed to study polymer glass formation[38–47]; the present modification is one of a family of potentials involving shortened bonds, which can yield improved crystallization resistance[48,49]. This potential has been employed in recent work probing the dynamics of polymer thin films during glass formation[50].

In simulations of dimers, each dimer is modeled as a rigid body such that the length of the intramolecular bond is fixed. We employ a polydisperse mixture of dimers, as follows. Each dimer simulation contains 5000 dimers. These dimers are divided into 25 equal groups and each group is assigned a distinct value of the LJ σ parameter. The values of σ reflect a uniform distribution over the range of 0.85 to 1.15, such that $\sigma_{ii}$ for group $i$ is given by the equation

$$\sigma_{ii} = 0.85 + \frac{0.3i}{25}. \quad (4)$$

In each case the fixed bond length within the dimer is also taken to be $\sigma_{ii}$. Nonbonded interactions between beads in distinct dimers of distinct size employ an arithmetic average range parameter:

$$\sigma_{ij} = \frac{\sigma_{ii} + \sigma_{jj}}{2}. \quad (5)$$

For each system we employ equilibrium starting configurations generated by our recently described PreSQ algorithm[51]. This algorithm consists of an iterative, adaptive, quench and anneal procedure that ensures equilibrations by requiring an annealing period of ten times the structural relaxation time (in this case the segmental reorientational relaxation time) prior to data collection.

We then perform simulations in the isoconfigurational ensemble. In this ensemble, a single starting configuration is copied many times and each copy is assigned a distinct Gaussian distribution of particle velocities. We specifically replicate each configuration and then collect dynamics for 1000 distinct trajectories emanating from each shared starting point. Dynamical properties are then measured for each such configuration and averaged over the full set, allowing normally ensemble properties to be computed on a single-particle basis.

We perform this procedure for two different temperatures for each system: one near the onset temperature $T_A$ below which dynamics first become non-Arrhenius, and one at a temperature at which the reorientational relaxation time is in the vicinity of $10^{3.5}$ LJ time units – the longest relaxation time at which this computationally expensive ensemble is presently tractable. In each case we run simulations long enough to considerably exceed the mean system structural relaxation time.

Simulations are performed in the Large-scale Atomic/Molecular Massively Parallel Simulator (LAMMPS) package[52] using a Verlet

integrator with a timestep of 0.005 $\tau_{LJ}$. We perform simulations in the NPT ensemble with a Nose-Hoover thermostat and barostat with damping parameters of 2 $\tau_{LJ}$ for polymers and thermostat of 2 $\tau_{LJ}$ and barostat of 5 $\tau_{LJ}$ in rigid body-dimers.

For every particle in each system, we compute mean translational and reorientational relaxation functions by averaging over all 1000 trajectories for that particle. We compute relaxation times based on several distinct pathways to ensure that our results are robust to the choice of relaxation function. Crucially, in all of these cases the calculation is based upon a single initial start time and is not averaged over multiple distinct time windows. As a consequence, the local relaxation behavior reported via all of these methods reflects the *first* relaxation event at that location and involves no temporal averaging.

We quantify translational dynamics three ways. First, we compute the self-part of the intermediate scattering function $F_s(q,t)$ at a wavenumber q = 7.07, comparable to the first peak in the structure factor, for translational dynamics. This is a common convention and reflects the lengthscale most closely associated with segmental relaxation. Second, we additionally probe behavior at a higher wavenumber of 12.5 to determine whether our qualitative findings are retained for relaxation at shorter length scales. Third, we compute a final translational relaxation time for each bead based on its mean-square displacement, defining the relaxation time as the time at which the mean square displacement for a particle equals 0.889, which is equal to $2\pi$ over the wavenumber of 7.07. We quantify reorientational dynamics two ways: via both the first and second Legendre polynomial bond autocorrelation function $C_1(t)$ and $C_2(t)$.

For all relaxation functions (all cases except the mean square displacement), we determine relaxation times and stretching exponents as follows. In order to enable a fit to the entire temperature range of relaxation, which generally involves both a ps-timescale relaxation and the segmental alpha process that is the focus this work, we fit the relaxation time data to a sum of two stretched exponential relaxation functions,

$$f(t) = A\exp\left[-\left(\frac{t}{\tau_{fast}}\right)^{\beta_{fast}}\right] + (1-A)\exp\left[-\left(\frac{t}{\tau_{slow}}\right)^{\beta}\right], \quad (6)$$

where $A$ is a constant that sets the fraction of relaxation attributable to the fast process. This type of two-term model has previously been shown to provide a good description for the entire relaxation function for bead-spring polymers[53]. The first of these terms fits the short time ps-relaxation. The second term fits the alpha process. The stretching exponent for the segmental α-relaxation is directly extracted as the value of β, which we refer to as $\beta^t$ and $\beta^r$ for translational and reorientational dynamics, respectively. We extract an overall relaxation time via the analytic expression for the zeroth moment of the entire equation (6), i.e.

$$\tau = \frac{A\tau_f}{\beta_{fast}}\Gamma\left(\frac{1}{\beta_{fast}}\right) + \frac{(1-A)\tau_{slow}}{\beta}\Gamma\left(\frac{1}{\beta}\right). \quad (7)$$

We again employ superscripts of "t" and "r" to denote translational and reorientational relaxation times, respectively. Since the fast relaxation involves a timescale $\tau_f$ that is always in the vicinity of 1 ps or less, when $\tau_{slow}$ is large, this converges to the relaxation time for the alpha process alone. However, for some fast particles, particularly at the higher temperatures probed, the alpha and beta processes are not well separated. In this case, this approach allows for unambiguous determination of relaxation times for these fast particles, taking account of the presence of a merged or partially-merged process.

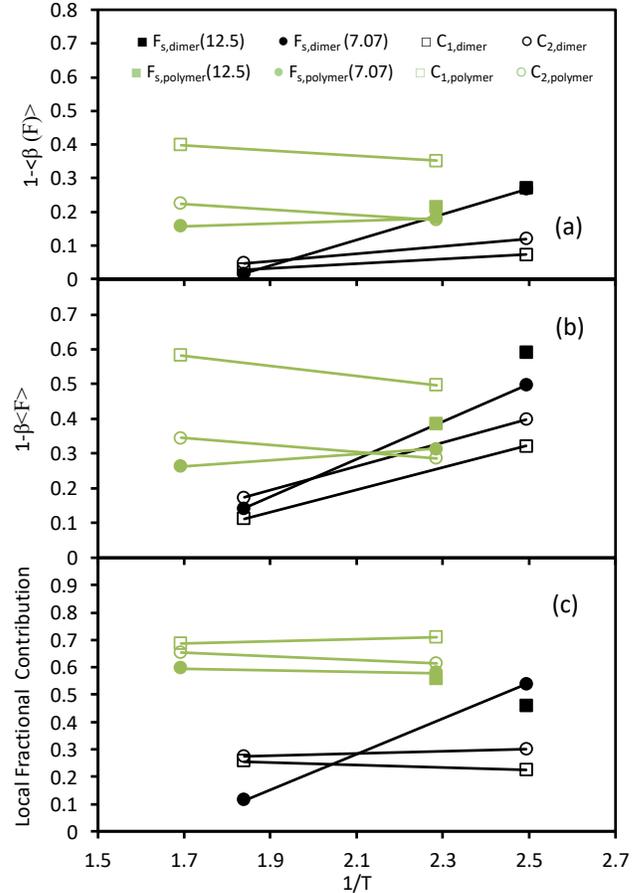

Figure 1. (a) Mean stretching exponent for local relaxation, (b) stretching exponent for mean system relaxation, and (c) fraction of *intrinsic* local contribution to stretching exponent, all vs inverse temperature. Symbols correspond to translational relaxation at q = 7.07 (filled circles), translational relaxation at q = 12.5 (filled squares), reorientational first Legendre polynomial relaxation (open squares), and reorientational second Legendre polynomial relaxation (open circles) for dimer (black symbols) and polymer (green symbols) systems. Lines are guides to the eye.

These bead-based simulations are performed in dimensionless LJ units. Throughout, we report temperatures in dimensionless LJ units, and time in LJ units $\tau_{LJ}$. There is no unique conversion of LJ temperature units to real units, but a conversion of 1000 K to one LJ temperature unit can yield results that are qualitatively sensible for many purposes. $\tau_{LJ}$ converts roughly to 1 ps.

## Results

### Contributions to stretching

We begin by probing the role of spatial averaging in relaxational stretching. To do so, we compute a stretching exponent two ways. First, we consider an average relaxation function *F* (either translational or reorientational) of all particles in the system. This is the mean system response that is commonly probed by experiment and most simulations. We then fit this to a stretched exponential form and extract a stretching exponent $\beta(<F>)$ for the mean system dynamics, which is the $\beta$ that would be measured by a typical 'mean system' approach. Alternately, we consider the relaxation function of each particle, taken separately and averaged over its 1000 trajectories from the same starting condition. We fit each particle's relaxation to a stretched exponential and determine the stretching exponent for that particle. We then take an average over the values of $\beta$ for all particles to determine the mean single-particle $<\beta(F)>$. In performing this analysis, we note that in all cases for the translational relaxation data at q=12.5 we include only the lowest temperature. For higher temperatures, relaxation is excessively contaminated by the fast beta relaxation at this wavenumber, such that it is not possible to obtain a meaningful value for the stretching exponent of the alpha relaxation process.

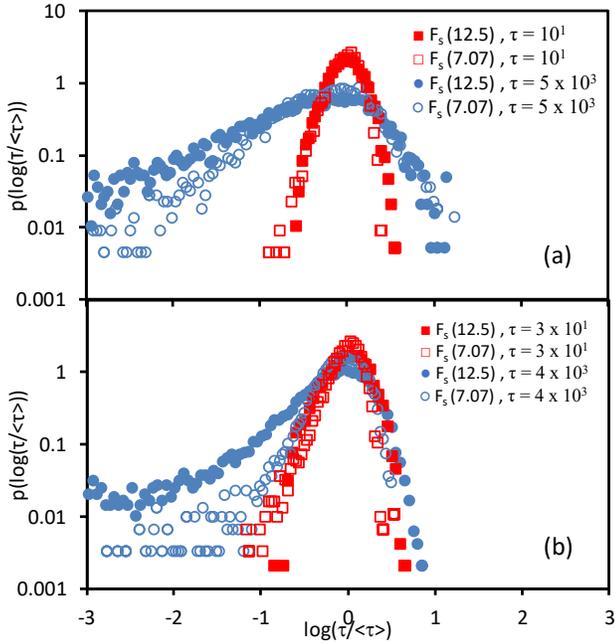

Figure 2. Distribution of local translational relaxation times at q = 12.5 (filled symbols) and q = 7.07 (open symbols) times for a Lennard-Jones dimer (a) and a bead spring polymer (b). Mean system relaxation times and wavenumber corresponding to each symbol are shown in the legends.

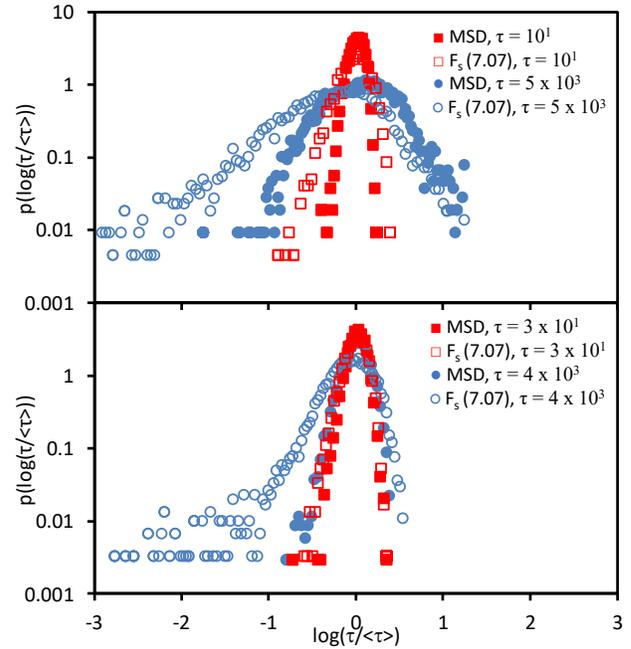

Figure 3. Distribution of local translational relaxation times obtained from the mean square displacement (filled symbols) and from $F_s(q,t)$ at q = 7.07 (open symbols) times for a Lennard-Jones dimer (a) and a bead spring polymer (b). In (a), red squares and blue circles correspond to temperatures with mean system reorientational relaxation times of $10^1$ $\tau_{LJ}$ and $5 \times 10^3$ $\tau_{LJ}$, respectively. In (b), these times are $3 \times 10^1$ $\tau_{LJ}$, and $4 \times 10^3$ $\tau_{LJ}$, respectively.

If averaging over spatial heterogeneity is the origin of relaxational stretching, then $<\beta(F)>$ should equal 1, since relaxation should be locally exponential, and $\beta(<F>)$ should be less than 1. As shown by Figure 1a, the pure spatial averaging hypothesis is insufficient to describe behavior. Instead, $<\beta(F)>$ is considerably less than one for both reorientational and translational dynamics, indicating that relaxation is locally stretched in a manner that cannot be explained by averaging over spatial dynamic heterogeneity. As shown in this figure, this holds true for translational dynamics both at the segmental length scale (q = 7.07) and more locally (q = 12.5) and for reorientational dynamics quantified two distinct ways. The finding that stretching does not emerge entirely from spatial average over dynamic heterogeneity thus appears to be quite robust. As above, we emphasize that the local values of β reflected in these data reflect the *first local* relaxation process and do not involve a temporal average over multiple observation windows ('start times'). These are thus intrinsic local stretching exponents not attributable to spatial or temporal averaging.

$\beta(<F>)$ for the mean system dynamics (Figure 1b) is lower still, indicating that spatial averaging further augments this local stretching. We can quantify the fraction of stretching that emerges from an *intrinsic* (non-temporally averaged) local origin (as opposed to spatial averaging) via the ratio $(1-<\beta(F)>)/(1-\beta(<F>))$. As shown by Figure 1c, this fraction is nonzero and can depend strongly on system and on choice of translational vs reorientational relaxation. In particular, local effects in the polymer play a relatively strong role over the entire temperature range probed, accounting for over half of the observed stretching in both translational and reorientational relaxation. This result is consistent with the relatively common observation of fairly temperature-insensitive stretching exponents in polymers as measured via rheology[54]. For dimer reorientation, on the other hand, local effects account for less than a third of the extent of stretching. The general finding that intrinsic stretching plays a weaker role in small-molecule reorientational stretching relative to polymer reorientational stretching appears to be consistent with Richert's synthesis of the experimental data[9]. This is also likely consistent with the observation above that chain connectivity in polymers contributes to intrinsic stretching[31].

Given that these models are locally similar at the scale of the dimer (or a pair of bonded repeat units in the chain), one might *very roughly* expect a similar level of intrinsic stretching not attributable to bond connectivity in the polymer as in the small molecule. This would suggest that around half of the polymer's observed intrinsic stretching might be attributable to chain connectivity effects and the other half to some more general glass physics. However, it is not clear that this is an entirely valid comparison. Further below we discuss other, more compelling, indicia of non-connectivity-driven intrinsic stretching in the polymeric system.

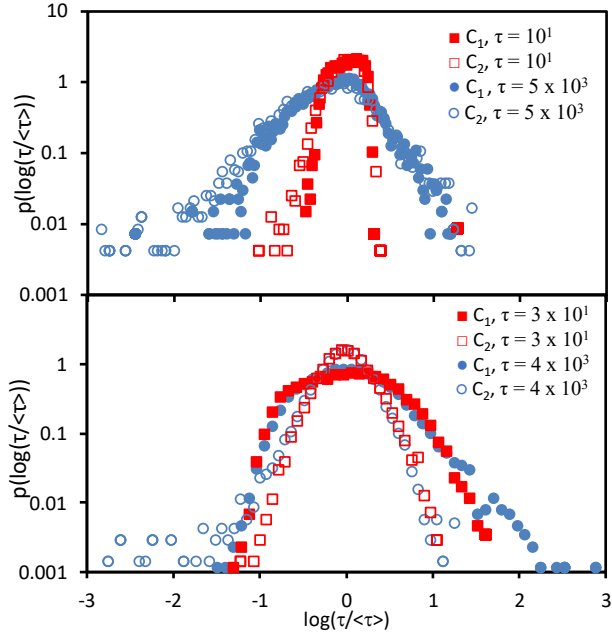

Figure 4. Distribution of local reorientational relaxation times computed from the first Legendre Polynomial reorientation function (filled symbols) and the second Legendre Polynomial reorientation function (open symbols) times for a Lennard-Jones dimer (a) and a bead spring polymer (b). In (a), red squares and blue circles correspond to temperatures with mean system reorientational relaxation times of $10^1$ $\tau_{LJ}$ and $5 \times 10^3$ $\tau_{LJ}$, respectively. In (b), these times are $3 \times 10^1$ $\tau_{LJ}$, and $4 \times 10^3$ $\tau_{LJ}$, respectively.

Aside from differences in magnitude, the temperature dependence of intrinsic stretching effects also differs between dimers and polymers. Dimer translation exhibits a strongly temperature dependent role for local effects, interpolating between a weak local contribution at high temperature and a polymer-like local contribution at low temperatures. Conversely, the overall degree of stretching, and the relative extent of local stretching, are less temperature dependent in the polymeric than the small molecule system. This may also be related to the fact that chain-connectivity effects can induce nonexponentiality in polymer segmental dynamics in a manner that is not associated with glass formation (and therefore should not grow on cooling)[31].

Below, we explore further the question of whether this behavior can be entirely attributed to chain connectivity effects in the polymer system. The data thus far, however, indicate that it is not generally safe to assume that the stretching exponent reports in some direct manner on the breadth of the underlying relaxation time distribution. More broadly, these results indicate that both spatial averaging *and* local nonexponentiality play important, system- and relaxation-function-dependent roles in the overall stretching behavior of glass-forming liquids.

### Origins of global and local stretching

What is the origin of the local and nonlocal contributions to stretching reported in Figure 1? To begin answering this question,

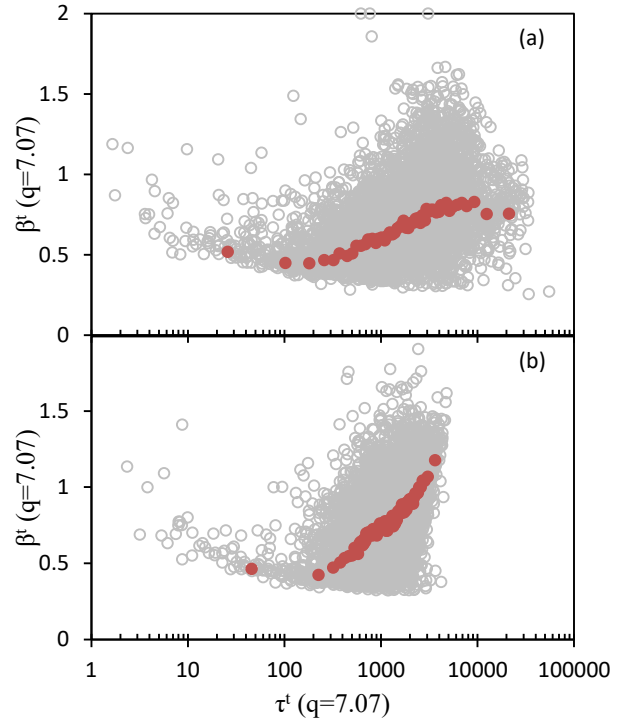

Figure 5. Local translational stretching exponent vs local translational relaxation time, both determined at at q = 7.07, at the lowest temperature simulated for: (a) dimer and (b) polymer. In both figures, grey open circles are values for individual particle relaxation functions, and red filled circles are values extracted from relaxations functions averaged over 100 particles of similar translational relaxation time.

we first quantify the underlying relaxation time distributions observed within the isoconfigurational ensemble. As can be seen in Figure 3, both the polymer and dimer exhibit a distribution of relaxation times for translational relaxation. The distributions observed at q = 12.5 and q = 7.07 are somewhat similar, with the exception that the higher wavenumber distribution exhibits a stronger tail to high mobilities. As shown in Figure 3, similar findings are also obtained if one instead extracts a relaxation time from the mean square displacement as described above. Qualitatively similar behavior is also observed for both measures of reorientational relaxation in Figure 4, although there are quantitative differences in the shape of the distributions that may provide a starting point of interest for future study.

In any of these cases, these distributions are *roughly* log-normal, albeit with an excess tail to the mobile particle side. This approximate log-normal behavior is consistent with prior work probing distributions of dynamical properties in glass-forming liquids[55], including experimental single-particle studies[29], and with theoretical work employing a Gaussian distribution of local relaxation barriers[22]. The overall form of the distribution including the excess mobile tail appears to be qualitatively consistent with results from single-particle experiments[28,29].

As can be seen in these data, the temperature dependences of these distributions differ considerably between the polymer and dimer. On cooling from $\sim T_A$ down to $\tau \sim 10^{3.5}$ $\tau_{LJ}$, the dimer relaxation time distributions broaden strongly, whereas the polymer relaxation time distributions broaden weakly. This difference explains why the nonlocal contribution to the stretching exponent grows appreciably on cooling for the dimer but not for the polymer – in the latter case the extent of dynamic heterogeneity does not grow as significantly on cooling over this temperature range.

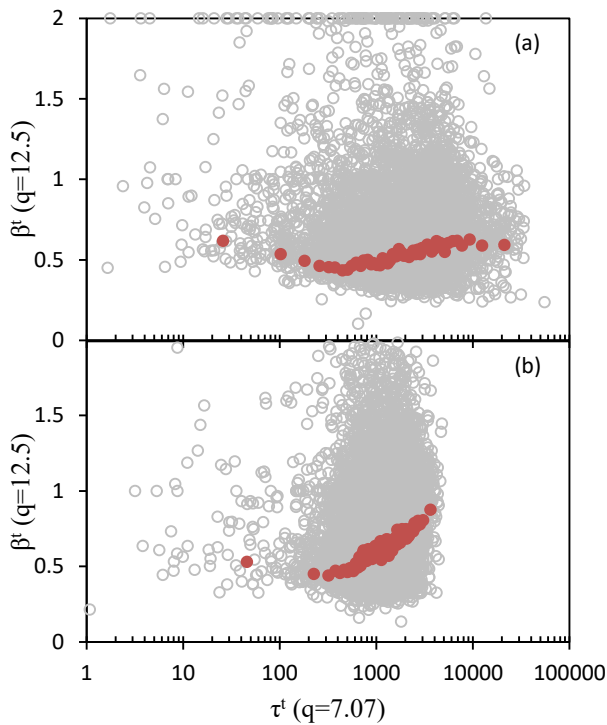

Figure 6. Local translational stretching exponent vs local translational relaxation time, both determined at q = 12.5, at the lowest temperature simulated for: (a) dimer and (b) polymer. In both figures, grey open circles are values for individual particle relaxation functions, and red filled circles are values extracted from relaxations functions averaged over 100 particles of similar translational relaxation time.

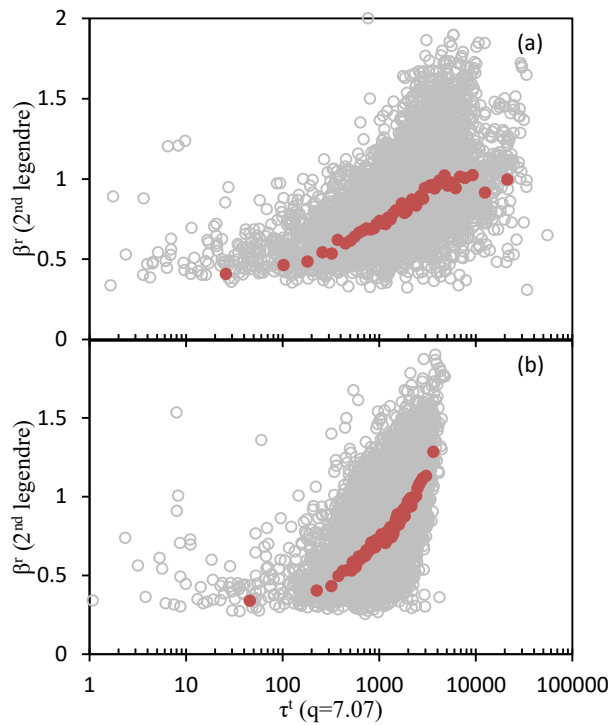

Figure 7. Local reorientational stretching exponent vs local translational relaxation time, both as obtained via the 2$^{nd}$ Legendre Polynomial, at the lowest temperature simulated for: (a) dimer and (b) polymer. In both figures, grey open circles are values for individual particle relaxation functions, and red filled circles are values extracted from relaxations functions averaged over 100 particles of similar translational relaxation time.

However, this heterogeneous origin is distinct from the intrinsic contributions to stretching reported in Figure 1. To better understand the origin of this phenomenon, we plot the value of the stretching exponent for each particle *vs* that particle's relaxation time. These results are shown in Figure 5, Figure 6, Figure 7, and Figure 8, for translational relaxation at q = 7.07, translational relaxation at q = 12.5, 2$^{nd}$ Legendre Polynomial reorientational relaxation, and 1$^{st}$ Legendre Polynomial Reorientational Relaxation, respectively. These data suggest that slower-relaxing particles, on average, tend to possess a higher value of β. However, even with an average over 1000 trajectories the underlying data are still noisy. To obtain further noise reduction, we sort the particles into groups of 100 based on their relaxation times. For each group, we average the relaxation function and then perform the stretched exponential fits. Because the particles' individual relaxation times are tightly clustered for each group, averaging effects are expected to play little role beyond noise reduction. A plot of β vs τ at this averaged level is incorporated in each of these figures and reveals a strong local correlation between β and τ for all relaxation functions probed: faster-relaxing regions tend to exhibit more stretching. This trend is evidently *qualitatively* consistent across multiple length scales and multiple types of relaxation.

The *quantitative magnitude* of the trend, on the other hand, is seen to depend upon relaxation function and choice of polymer vs dimer. It is generally stronger for the polymer than the dimer system, although it is clearly present in both. Remarkably, in a number of these cases, the slowest-relaxing regions exhibit values of β on average approaching and even exceeding one, indicating that their relaxation is compressed. This trend can visually be seen in representative curves for translational and reorientational relaxation functions shown in Figure 9 and Figure 10. What is the origin of this behavior?

Notably, a parallel result was reported in recent single-molecule fluorescent probe experiments, which reported that particles "with high β are concentrated in the slower part of the distribution".[28] In that work, this was attributed to temporal averaging over more relaxation periods in fast-relaxing particles than in slow-relaxing particles, such that the faster relaxing particles probed more of the distribution of relaxation times in the samples. In contrast, our results can firmly exclude this proposed explanation. Whereas single molecule experiments average over multiple relaxation times of each particle, here the value of β reported for each particle reflects its *first* relaxation event at the start of the simulation, averaged over the isoconfigurational ensemble of multiple initial velocities. Thus, *no* temporal averaging is present in these simulations, and temporal averaging is thus not a possible explanation for this observation within our data.

We cannot *firmly* exclude the possibility that the mechanism for this phenomenon may be different here than in the experimental systems probed by Paeng et al. However, these data indicate that temporal averaging is certainly not *required* to explain apparent local stretching; consideration of potential alternate origins of local (genuinely intrinsic) stretching is in order. As discussed above, our results also appear to be in reasonable qualitative agreement with the experimental results.

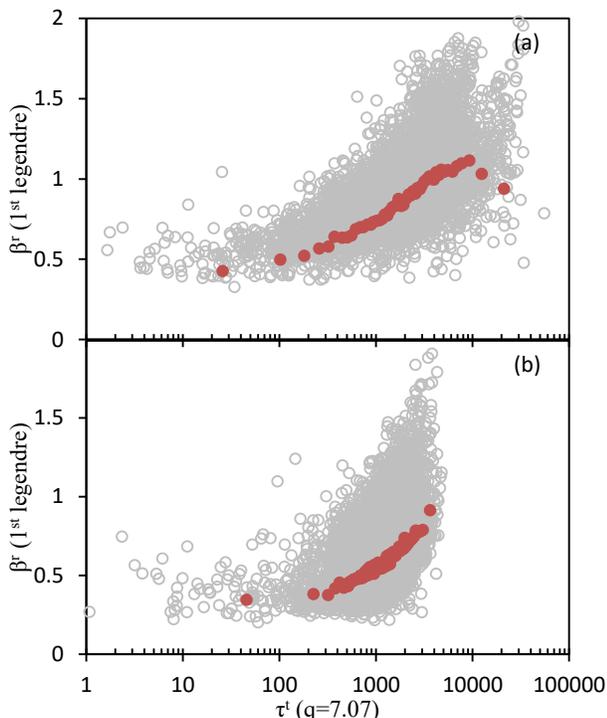

Figure 8. Local reorientational stretching exponent vs local translational relaxation time, both as obtained via the 2$^{nd}$ Legendre Polynomial, at the lowest temperature simulated for: (a) dimer and (b) polymer. In both figures, grey open circles are values for individual particle relaxation functions, and red filled circles are values extracted from relaxations functions averaged over 100 particles of similar translational relaxation time.

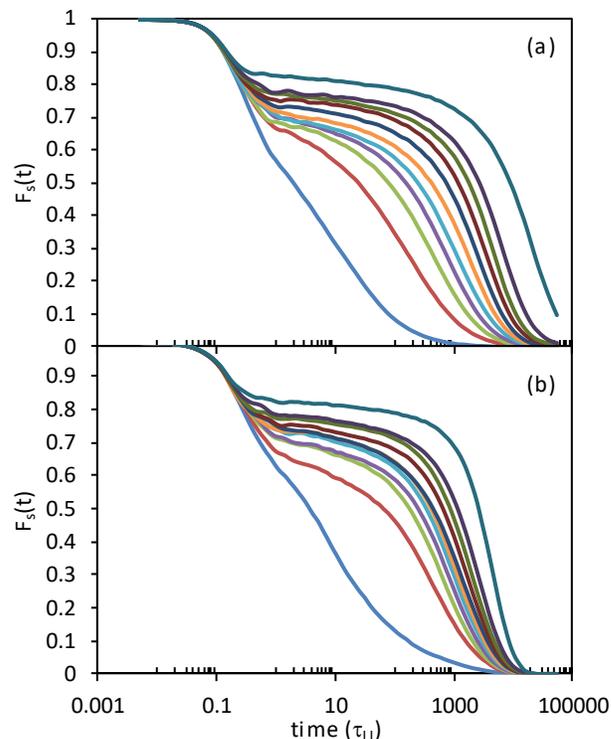

Figure 9. Self-part of the intermediate scattering function at q = 7.07 vs log time for dimer (a) and polymer (b). Each curve represents an average over 100 particles of like relaxation times, with curves from top to bottom reflecting particles at 100th, 90th, 80th, 70th, 60th, 50th, 40th, 30th, 20th, 10th, and 2th percentile respectively.

Moreover, the presence of this same trend towards more compressed relaxation in slower-relaxing particles in small molecule *and* polymer systems indicates that local stretching in the polymer cannot be attributed entirely to the chain connectivity effects discussed in the introduction. Evidently some mechanism not deriving from chain connectivity and not requiring spatial averaging is required to explain this behavior.

Given these facts, how can this trend in β vs τ be understood? Notably, this behavior – more stretched relaxation in faster relaxing regions and more compressed relaxation in slower relaxing regions – phenomenologically parallels the *mean system relaxation* of glasses during aging[56]. The pioneering work of Kovacs[57] established that under-dense and over-dense glasses exhibit an asymmetry of approach to equilibrium at equal aging temperature, with under-dense (fast-relaxing) glasses exhibiting more stretched (lower-β) relaxation, and over-dense (slow-relaxing) glasses exhibiting relatively more compressed (or less stretched, higher-β) relaxation. The usual explanation for this phenomenon is that over-dense glasses tend to become less dense, and thus faster relaxing, as they age, leading to an enhancement in mobility with time (autoacceleration). This naturally leads to compression of the relaxation process relative to equilibrium, with the inverse scenario naturally producing stretching in initially under-dense glasses[56]. Notably, this does not always require averaging over multiple local aging times; rather the degree of mobility of particles can develop over a single effective aging time of the material.

It is possible that an analogous phenomenon is responsible for the asymmetry of local equilibrium relaxation in these equilibrium liquids. This can occur if relaxation of slow particles tends to enhance their mobility (and vice versa in fast regions) *over a single local relaxation time of the selected particle*. This does not involve temporal or spatial averaging in any substantive sense, since it reflects a single particle over a single local relaxation time – rather it is a reflection of the fact that even a single relaxation event tends to change a given particle's mobility. Put another way, this does not require that ergodicity is restored over the timescale of observation – indeed, a single local relaxation time is almost certainly insufficient to recover ergodicity via full exchange of fast and slow regions. Instead, within this scenario, intrinsic local stretching and compression in glass-forming liquids is a *consequence* of dynamical heterogeneity, but it is **not** a result of spatial or temporal averaging over this heterogeneity. This suggests that dynamically heterogeneous systems are both intrinsically locally heterogeneous (because single relaxation events alter mobility) *and* introduce spatial averaging – two distinct origins of heterogeneity.

If this mechanism is correct, there should be a signature of locally 'under-dense' and 'over-dense' states in the fast-relaxing and slow-relaxing domains, respectively. One useful measure of local packing, as reflected in dynamics themselves, is the Debye Waller factor $<u^2>$. $<u^2>$ measures the segmental or molecular rattle space in a transient caging regime and is commonly defined in simulation as $<u^2> = <r^2(t = 1ps)>$[53,58]. In general, lower values of $<u^2>$ are a signature of dynamically tighter local packing and vice-versa[59].

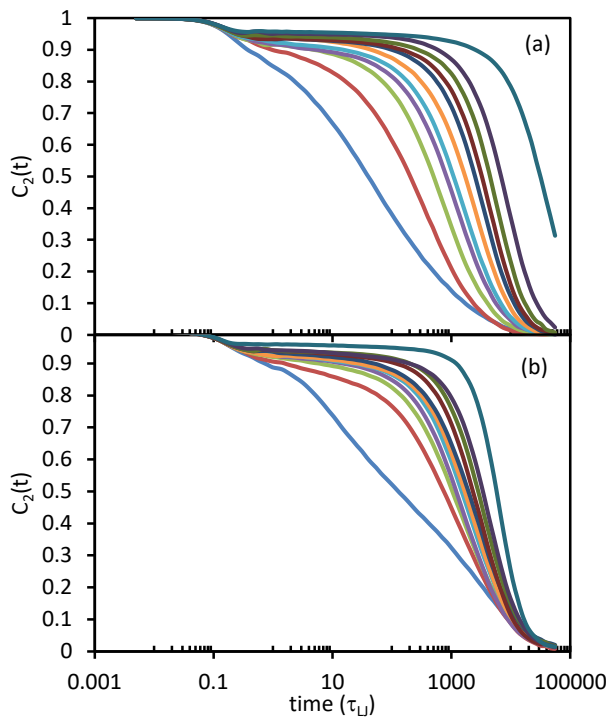

Figure 10. Second Legendre Polynomial reorientational relaxation times vs log time for dimer (a) and polymer (b). Each curve represents an average over 100 particles of like relaxation times, with curves from top to bottom reflecting particles at 100th, 90th, 80th, 70th, 60th, 50th, 40th, 30th, 20th, 10th, and 2th percentile respectively.

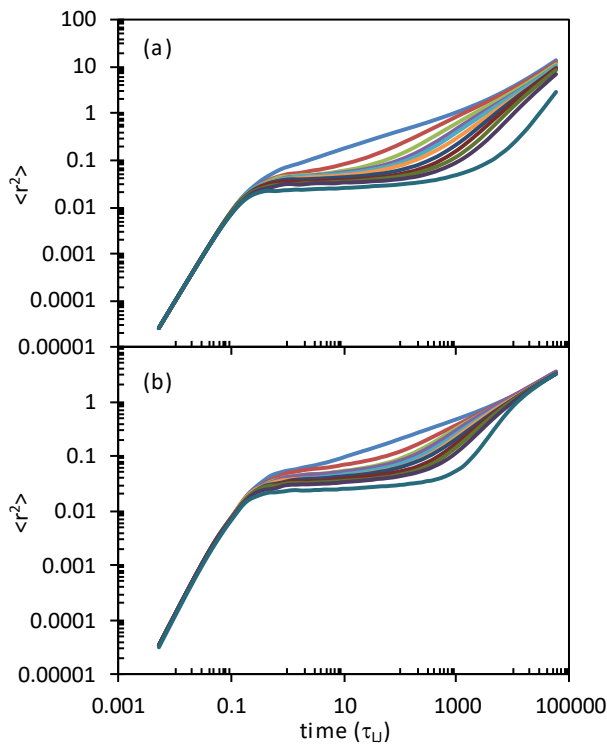

Figure 11. Mean square displacement vs time for dimer (a) and polymer (b). Each curve represents an average over 100 particles of like translational relaxation times, with curves from bottom to reflectiing particles at 100th, 90th, 80th, 70th, 60th, 50th, 40th, 30th, 20th, 10th, and 2nd percentile respectively.

We thus continue by quantifying the local mean-square-displacement in these systems for groups of beads of similar relaxation time. As shown by Figure 11, local $<r^2(t)>$ curves exhibit a spectrum of behavior comparable to $F_s(q,t)$. Slower-relaxing particles generally exhibit more pronounced caging, followed by a more abrupt cage escape, relative to faster-relaxing particles. As shown by Figure 12, $<u^2>$ values extracted from these curves are strongly predictive of stretching exponents: in general, particles that exhibit tighter caging also exhibit more compressed relaxation, while particles that exhibit looser caging exhibit more stretched relaxation.

This finding is consistent with the connection to aging of 'over-dense' and 'under-dense' glasses proposed above: just as occurs out of equilibrium at a mean system level, locally in-equilibrium domains with tighter packing relax more slowly and in a relatively more compressed manner, whereas domains with looser packing relax more quickly and in a more stretched manner. Conversely, relaxation in below-average $<u^2>$ domains must tend to increase their local $<u^2>$ and vice versa. This would be entirely unsurprising over many multiples of the relaxation time, since it is required to restore ergodicity. The finding, here, that this may play a significant role over single local relaxation times in both slow and mobile particles is relatively unexpected.

Finally, we note that these findings accord with several prior works that amplify this perspective. First, prior work by Leporini and coworkers have identified a strong correlation between local $<u^2>$ and local relaxation times in thin films[60]. These findings indicate that local rattle space plays an important causal role in controlling local relaxation. These findings also naturally connect to work of Riggleman et al. probing local elastic modulus distributions in glasses and glass-forming liquids[55]. That study suggested the presence of a broad heterogeneous distribution of local elastic moduli, including a population of regions with *negative* local elastic modulus. That work also suggested an intimate connection between this modulus distribution and the distribution of local Debye-Waller factors. It thus seems likely that local stretching and compression behavior is also connected to local variation in elastic modulus. This possibility warrants additional investigation.

## Conclusion

These simulations in the isoconfigurational ensemble suggest that stretched relaxation in glass-forming liquids is not purely a consequence of spatial averaging over a distribution of locally exponential processes. Instead, local relaxation, *even at a single particle level without time averaging*, exhibits a spectrum of non-exponential relaxation behavior in our simulated systems. Results indicate that fast-relaxing particles tend to exhibit locally stretched relaxation, whereas slow-relaxing particles tend to exhibit locally relatively compressed (or less stretched) relaxation. In the system average, the more stretched relaxation tends to dominate in the systems we have simulated, such that the mean local relaxation process is stretched. Spatial averaging over a distribution of relaxation times then further increases the degree of stretching as measured from the mean system relaxation function, which is the quantity normally probed in experiment and most simulations.

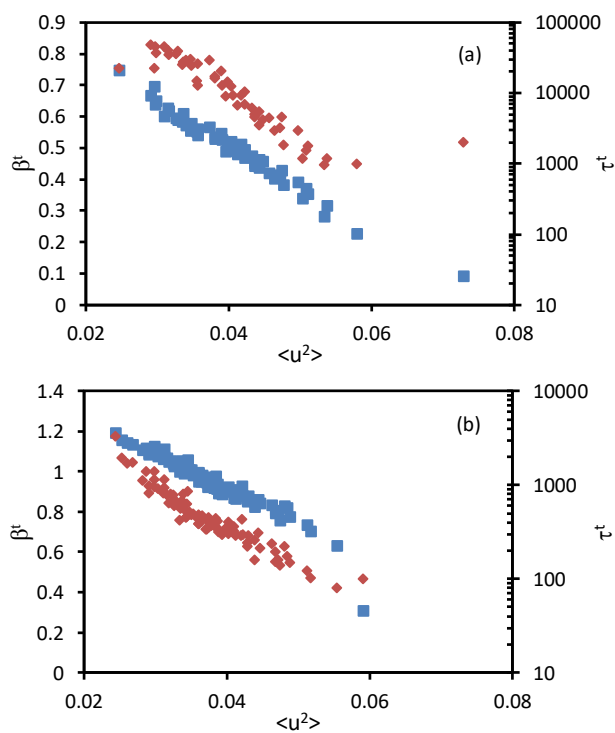

Figure 12. Translational stretching exponents (left vertical axis, red diamonds) and relaxation times (right vertical axis, blue squares) vs Debye-Waller factors for dimer (a) and polymer (b). Each point corresponds to a datum extracted from relaxations functions averaged over 100 particles of similar translational relaxation time.

The observation of an inverse correlation between local relaxation time and local non-exponentiality is consistent with results from single particle experiments[28]. Because that experimental work unavoidably involved integration over many local (and mean system) relaxation times, there it was suggested that temporal averaging was responsible for this effect via slow/fast exchange and ergodicity restoration. Here we can exclude this as an explanation for this behavior in our simulated systems, because they involve no temporal averaging, essentially reflecting the first local relaxation event (averaged over the isoconfigurational ensemble).

We show that these qualitative findings apply for both a model small molecule (a bead-spring dimer) and a model bead-spring polymer. In the latter of these two cases, chain connectivity effects may play a quantitative roll in the observed results (indeed, there are significant quantitative differences between the polymer and small molecule systems), but the shared qualitative phenomenology between the systems indicates that a polymer connectivity effect is not the underlying origin of the behavior reported here. Moreover, while the mean system stretching behavior can differ considerably between translational and reorientational relaxation functions, the findings above hold for both translational and reorientational relaxation[23]. The results for translational dynamics are also shown to be qualitatively robust to choice of q over the range 7.07 to 12.5, despite the fact that the extent of mean system relaxational stretching can var considerably over this range[61].

The basic phenomenonology of stretching at a local level appears to represent an in-equilibrium analogue of the dynamics of out of equilibrium glasses during aging. Single relaxation events of fast (loosely packed) particles involve a reduction in mobility over a single relaxation time; single relaxation events of slow (tightly packed) particles involve an enhancement in mobility over a single relaxation time. This occurs over timescales logically insufficient to allow restoration of ergodicity via full exchange of fast and slow regions, since this exchange cannot reasonably take place over a single particle relaxation time. It thus cannot be attributed to spatial and temporal averaging, and the resulting stretching instead appears to be an intrinsic, local feature of dynamically heterogeneous systems that does not emerge via a spatial or temporal averaging mechanism.

From a practical standpoint, these results may have significant implications for interpretation of stretching exponents obtained from experimental methods such as dielectric spectroscopy. In our simulated systems, the stretching exponent, and its temperature dependence, cannot not be interpreted as a literal measure of dynamic heterogeneity via a spatial averaging scenario. We find that $\beta$ involves strong intrinsic local nonexponentialities that emerge from heterogeneity via a distinct mechanism not involving spatial or temporal averaging, and this can play a major role in $\beta$'s temperature dependence. The relative contribution of these local nonexponentialities, and their temperature dependence, depend on system and relaxation function. *These results indicate that, while $\beta$ may correlate with the extent of local heterogeneity in some systems, one cannot take beta to be a quantitative measure of the extent of dynamic heterogeneity in general.* It also appears that in general, there is not a 'single' stretching exponent for a given system – different relaxation functions can exhibit different degrees of stretching.

Finally, we also note that these findings may have implications for the understanding of the glass transition. The polymeric and small molecule glass-formers exhibit similarly non-Arrhenius dynamics over this relaxation time range, yet the extents to which their relaxation time distributions broaden on cooling differ considerably. Evidently, broadening of the relaxation time distribution upon cooling is not a quantitatively universal aspect of supercooled liquid dynamics.

## Acknowledgments

This material is based on work supported by the National Science Foundation NSF CAREER Award, under grant number DMR-1554920.

## Data Availability Statement

The data that support the findings of this study are available from the corresponding author upon reasonable request.

## References

[1] A. Cavagna, Phys. Rep.-Rev. Sec. Phys. Lett. **476**, 51 (2009).
[2] M.D. Ediger, Annual Review of Physical Chemistry **51**, 99 (2000).

[3] M.T. Cicerone, P.A. Wagner, and M.D. Ediger, J. Phys. Chem. B **101**, 8727 (1997).
[4] G.S. Matharoo, M.S.G. Razul, and P.H. Poole, Phys. Rev. E **74**, 050502 (2006).
[5] F.H.M. Zetterling, M. Dzugutov, and S.I. Simdyankin, Journal of Non-Crystalline Solids **293**, 39 (2001).
[6] H. Tanaka, T. Kawasaki, H. Shintani, and K. Watanabe, Nature Materials **9**, 324 (2010).
[7] J.A. Rodriguez Fris, E.R. Weeks, F. Sciortino, and G.A. Appignanesi, Phys. Rev. E **97**, 060601 (2018).
[8] W.K. Kegel and and A. van Blaaderen, Science **287**, 290 (2000).
[9] R. Richert, J. Phys.: Condens. Matter **14**, R703 (2002).
[10] E.V. Russell and N.E. Israeloff, Nature **408**, 695 (2000).
[11] A. Arbe, J. Colmenero, M. Monkenbusch, and D. Richter, Phys. Rev. Lett. **81**, 590 (1998).
[12] M.D. Ediger, C.A. Angell, and S.R. Nagel, Journal of Physical Chemistry **100**, 13200 (1996).
[13] J. Qian, R. Hentschke, and A. Heuer, J. Chem. Phys. **110**, 4514 (1999).
[14] B. Shang, J. Rottler, P. Guan, and J.-L. Barrat, Phys. Rev. Lett. **122**, 105501 (2019).
[15] J. Baschnagel and F. Varnik, Journal of Physics: Condensed Matter **17**, R851 (2005).
[16] C. Cammarota and G. Biroli, J. Chem. Phys. **138**, 12A547 (2013).
[17] W. Kob, S. Roldán-Vargas, and L. Berthier, Nature Physics **8**, 164 (2012).
[18] G.M. Hocky, L. Berthier, W. Kob, and D.R. Reichman, Phys. Rev. E **89**, 052311 (2014).
[19] L. Berthier and W. Kob, Phys. Rev. E **85**, 011102 (2012).
[20] Y.-W. Li, Y.-L. Zhu, and Z.-Y. Sun, J. Chem. Phys. **142**, 124507 (2015).
[21] S.K. Kumar, G. Szamel, and J.F. Douglas, Journal of Chemical Physics **124**, 214501 (2006).
[22] K.S. Schweizer and E.J. Saltzman, J. Phys. Chem. B **108**, 19729 (2004).
[23] J.-H. Hung, J.H. Mangalara, and D.S. Simmons, Macromolecules **51**, 2887 (2018).
[24] A.P. Sokolov and K.S. Schweizer, Phys. Rev. Lett. **102**, 248301 (2009).
[25] Y. Cheng, J. Yang, J.-H. Hung, T.K. Patra, and D.S. Simmons, Macromolecules **51**, 6630 (2018).
[26] K.L. Ngai, D.J. Plazek, and C.M. Roland, Phys. Rev. Lett. **103**, 159801 (2009).
[27] K. Paeng and L. J. Kaufman, Chemical Society Reviews **43**, 977 (2014).
[28] K. Paeng, H. Park, D.T. Hoang, and L.J. Kaufman, Proc Natl Acad Sci U S A **112**, 4952 (2015).
[29] K. Paeng and L.J. Kaufman, Macromolecules **49**, 2876 (2016).
[30] D.T. Hoang, K. Paeng, H. Park, L.M. Leone, and L.J. Kaufman, Anal. Chem. **86**, 9322 (2014).
[31] A. Heuer and K. Okun, J. Chem. Phys. **106**, 6176 (1997).
[32] N. Giovambattista, S.V. Buldyrev, F.W. Starr, and H.E. Stanley, Physical Review Letters **90**, 085506 (2003).
[33] J. Fan, H. Emamy, A. Chremos, J.F. Douglas, and F.W. Starr, J. Chem. Phys. **152**, 054904 (2020).
[34] B.A.P. Betancourt, J.F. Douglas, and F.W. Starr, Journal of Chemical Physics **140**, 204509 (2014).
[35] F.W. Starr, J.F. Douglas, and S. Sastry, The Journal of Chemical Physics **138**, 12A541 (2013).
[36] A. Widmer-Cooper and P. Harrowell, Phys. Rev. Lett. **96**, 185701 (2006).
[37] K. Kremer and G.S. Grest, The Journal of Chemical Physics **92**, 5057 (1990).
[38] F. Varnik, J. Baschnagel, and K. Binder, Phys. Rev. E **65**, 021507 (2002).
[39] J. Buchholz, W. Paul, F. Varnik, and K. Binder, The Journal of Chemical Physics **117**, 7364 (2002).
[40] F.W. Starr and J.F. Douglas, Phys. Rev. Lett. **106**, 115702 (2011).
[41] Paul Z. Hanakata, Jack F. Douglas, and Francis W. Starr, Journal of Chemical Physics **137**, 244901 (2012).
[42] J. Baschnagel and F. Varnik, Journal of Physics: Condensed Matter **17**, R851 (2005).
[43] J.-L. Barrat, J. Baschnagel, and A. Lyulin, Soft Matter **6**, 3430 (2010).
[44] J.H. Mangalara and D.S. Simmons, ACS Macro Lett. **4**, 1134 (2015).
[45] J.H. Mangalara, M.D. Marvin, N.R. Wiener, M.E. Mackura, and D.S. Simmons, The Journal of Chemical Physics **146**, 104902 (2017).
[46] M. Chowdhury, Y. Guo, Y. Wang, W.L. Merling, J.H. Mangalara, D.S. Simmons, and R.D. Priestley, J. Phys. Chem. Lett. 1229 (2017).
[47] D. Ruan and D.S. Simmons, Macromolecules **48**, 2313 (2015).
[48] M.E. Mackura and D.S. Simmons, Journal of Polymer Science Part B: Polymer Physics **52**, 134 (2014).
[49] J.H. Mangalara, M.E. Mackura, M.D. Marvin, and D.S. Simmons, The Journal of Chemical Physics **146**, 203316 (2017).
[50] D. Diaz-Vela, J.-H. Hung, and D.S. Simmons, ACS Macro Lett. 1295 (2018).
[51] J.-H. Hung, T. Patra, and D. Simmons, in (2016), p. E33.002.
[52] S.J. Plimpton, J. Comp. Phys. **117**, 1 (1995).
[53] D.S. Simmons and J.F. Douglas, Soft Matter **7**, 11010 (2011).
[54] R. Inoue, T. Kanaya, K. Nishida, I. Tsukushi, M.T.F. Telling, B.J. Gabrys, M. Tyagi, C. Soles, and W. -l. Wu, Phys. Rev. E **80**, 031802 (2009).
[55] R.A. Riggleman, J.F. Douglas, and J.J. de Pablo, Soft Matter **6**, 292 (2010).
[56] G.B. McKenna and S.L. Simon, Macromolecules **50**, 6333 (2017).
[57] A.J. Kovacs, in *Fortschritte Der Hochpolymeren-Forschung* (Springer Berlin Heidelberg, 1964), pp. 394–507.
[58] L. Larini, A. Ottochian, C. De Michele, and D. Leporini, Nature Physics **4**, 42 (2008).
[59] D.S. Simmons, M.T. Cicerone, Q. Zhong, M. Tyagi, and J.F. Douglas, Soft Matter **8**, 11455 (2012).
[60] M. Becchi, A. Giuntoli, and D. Leporini, Soft Matter **14**, 8814 (2018).
[61] C. Rehwald and A. Heuer, Phys. Rev. E **86**, 051504 (2012).